\documentclass[aps,prc,onecolumn,notitlepage,superscriptaddress,showkeys,nofootinbib,floatfix,10pt]{revtex4-1}
\usepackage{graphicx,amsmath,verbatim,color}
\usepackage[left=2cm,right=2cm,top=2cm,bottom=2cm,headheight=1cm]{geometry}
\usepackage[charter]{mathdesign}
\definecolor{darkblue}{RGB}{46,48,146}
\usepackage{ifpdf}
\ifpdf
\usepackage[
	pdfstartview=FitH,
	pdfpagemode=UseNone,
  colorlinks=true,
	linkcolor=darkblue,
	filecolor=darkblue,
	urlcolor=darkblue,
	citecolor=darkblue,
	bookmarks=false,
	pdftex,
	plainpages=false,
	hypertexnames=false,
	pdfpagelabels=true,
	hyperindex=true]{hyperref}
\DeclareGraphicsExtensions{.pdf}
\else
\DeclareGraphicsExtensions{.eps}
\fi

\begin{document}

\title{Quadrupole Moments of $^\textbf{29}$Mg and $^\textbf{33}$Mg}

\author{\mbox{Deyan~Todorov~Yordanov}}
\email[]{Deyan.Yordanov@cern.ch}
\affiliation{\mbox{Institut de Physique Nucl\'eaire, CNRS-IN2P3, Universit\'e Paris-Sud, Universit\'e Paris-Saclay, F-91406 Orsay, France}}
\affiliation{\mbox{Max-Planck-Institut f\"ur Kernphysik, Saupfercheckweg 1, D-69117 Heidelberg, Germany}}
\affiliation{\mbox{CERN European Organization for Nuclear Research, Physics Department, CH-1211 Geneva 23, Switzerland}}

\author{\mbox{Magdalena~Kowalska}}
\email[]{Magdalena.Kowalska@cern.ch}
\affiliation{\mbox{CERN European Organization for Nuclear Research, Physics Department, CH-1211 Geneva 23, Switzerland}}

\author{\mbox{Klaus~Blaum}}
\email[]{Klaus.Blaum@mpi-hd.mpg.de}
\affiliation{\mbox{Max-Planck-Institut f\"ur Kernphysik, Saupfercheckweg 1, D-69117 Heidelberg, Germany}}

\author{\mbox{Marieke~De~Rydt}}
\affiliation{\mbox{Instituut voor Kern- en Stralingsfysica, KU Leuven, Celestijnenlaan 200D, BE-3001 Leuven, Belgium}}

\author{\mbox{Kieran~T.~Flanagan}}
\email[]{Kieran.Flanagan-2@manchester.ac.uk}
\affiliation{School of Physics and Astronomy, The University of Manchester, Manchester, M13 9PL, United Kingdom}

\author{\mbox{Pieter~Himpe}}
\affiliation{\mbox{Instituut voor Kern- en Stralingsfysica, KU Leuven, Celestijnenlaan 200D, BE-3001 Leuven, Belgium}}

\author{\mbox{Peter~Lievens}}
\email[]{Peter.Lievens@kuleuven.be}
\affiliation{\mbox{Laboratorium voor Vaste-Stoffysica en Magnetisme, KU Leuven, Celestijnenlaan 200D, BE-3001 Leuven, Belgium}}

\author{\mbox{Stephen~Mallion}}
\affiliation{\mbox{Instituut voor Kern- en Stralingsfysica, KU Leuven, Celestijnenlaan 200D, BE-3001 Leuven, Belgium}}

\author{\mbox{Rainer~Neugart}}
\email[]{Rainer.Neugart@uni-mainz.de}
\affiliation{\mbox{Max-Planck-Institut f\"ur Kernphysik, Saupfercheckweg 1, D-69117 Heidelberg, Germany}}
\affiliation{\mbox{Institut f\"ur Kernchemie, Universit\"at Mainz, D-55128 Mainz, Germany}}

\author{\mbox{Gerda~Neyens}}
\email[]{Gerda.Neyens@kuleuven.be}
\affiliation{\mbox{Instituut voor Kern- en Stralingsfysica, KU Leuven, Celestijnenlaan 200D, BE-3001 Leuven, Belgium}}

\author{\mbox{Nele~Vermeulen}}
\affiliation{\mbox{Instituut voor Kern- en Stralingsfysica, KU Leuven, Celestijnenlaan 200D, BE-3001 Leuven, Belgium}}

\author{\mbox{Henry Stroke}}
\email[]{Henry.Stroke@nyu.edu}
\affiliation{\mbox{Department of Physics, New York University, New York, NY 10003, USA}}

\date{\today}

\begin{abstract}

The quadrupole moments of $^{29}$Mg and $^{33}$Mg have been constrained by collinear laser spectroscopy at CERN-ISOLDE. The values are consistent with shell-model predictions, thus supporting the current understanding of light nuclei associated with the ``island of inversion''.

\end{abstract}


\maketitle

\section{Introduction}

The deformation in light nuclei in proximity of $N=20$ has been studied extensively since the 1970's when it was first discovered \cite{Thibault,Huber,Detraz}. The magnesium isotopes in particular attracted considerable attention, including measurements by collinear laser spectroscopy \cite{Neyens,Kowalska,DTYordanov,Yordanov}, in an effort to uncover the boundaries of this island of inversion, known as such due to the apparent inversion of states in the relevant shell-model description. As of now, it is well understood that $^{29}$Mg \cite{Kowalska} follows the expected level ordering in $sd$-shell nuclei, while $^{33}$Mg \cite{DTYordanov} has a ground state determined by particle-hole excitations across the $N=20$ shell gap. There has been much debate \cite{Tripathi,DTY,Kanungo} as to what is the exact number of particle-hole excitations involved in $^{33}$Mg and the associated parity assignment. The experimental evidence supports a $3/2^-$ configuration, as outlined in the critical evaluation by Neyens \cite{GNeyens}.

\section{Experiment}

The quadrupole moments presented here are not derived from dedicated measurements as they could be considered a byproduct from the $\beta$-asymmetry detection and NMR work on the two cases \cite{Kowalska,DTYordanov}. Collinear laser spectroscopy was carried out at CERN-ISOLDE with the experimental setup discussed in the original publications. The ions of $^\text{29, 33}$Mg were excited in the transition $3s\;^2S_{1/2}\rightarrow 3p\;^2P_{3/2}$ at $280$~nm \cite{Kaufman} which provides quadrupole interaction in the excited state. At the time, the hyperfine structure was utilized primarily as a tool for aiding the NMR measurements by observing the amount of nuclear orientation produced by optical pumping and as a signature of the signs of the electromagnetic moments. Consequently, the quadrupole splitting was not discussed in Refs.~\cite{Kowalska,DTYordanov}. This was partly justified by the fact that the hyperfine structure in the $3p\;^2P_{3/2}$ state is not resolved, thus suffering a power-dependent cross-pumping effect that needed to be understood. It was later shown \cite{Yordanov} that $\beta$-asymmetry measurements of the atomic hyperfine structure could be reproduced quantitatively with the formalism outlined Refs.~\cite{Keim,Kowalska}. We are therefore making at attempt to evaluate the quadrupole moments of $^\text{29, 33}$Mg from the existing and limited data by using the realistic polarization fit function described in the aforementioned work.

\section{Results}

\begin{figure}[t]
	\begin{center}
		\begin{minipage}[b]{0.45\textwidth}
			\includegraphics[angle=0,width=\linewidth]{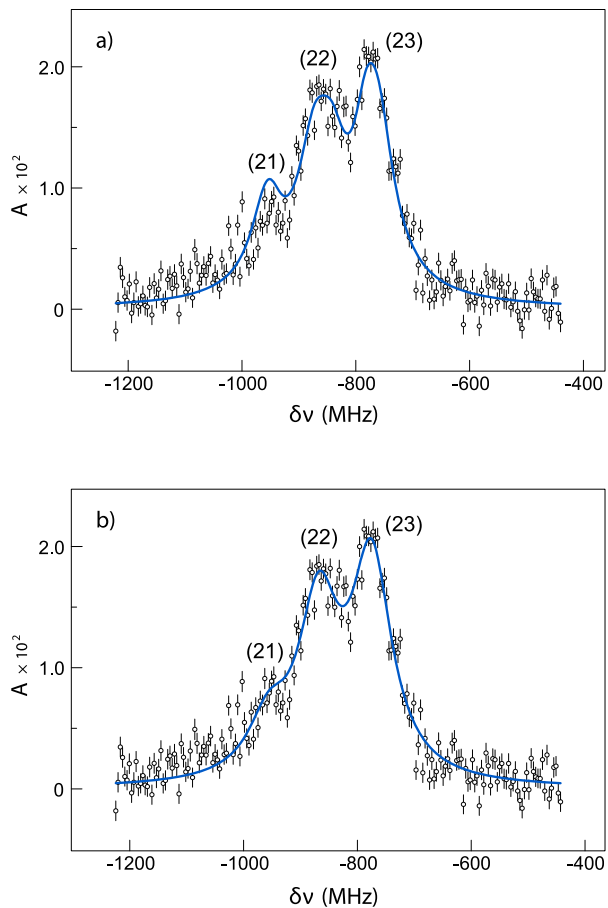}
			\caption{$\beta$-asymmetry spectrum of $\sigma^+$ polarized $^{29}$Mg\,\textsc{ii} ($I^\pi=3/2^+$) fitted with: (a) The realistic polarization function discussed in the text; (b) Lorentzian profiles. The corresponding $F$ quantum numbers are indicated above the individual peaks. The frequency range covers the lower-energy part of the spectrum associated with excitations from the $F=2$ hyperfine member of the $3s\;^2S_{1/2}$ state.}
			\label{fig1}
		\end{minipage}
	\end{center}
\end{figure}

\begin{figure}[t]
	\begin{center}
		\begin{minipage}[b]{0.45\textwidth}
			\includegraphics[angle=0,width=\linewidth]{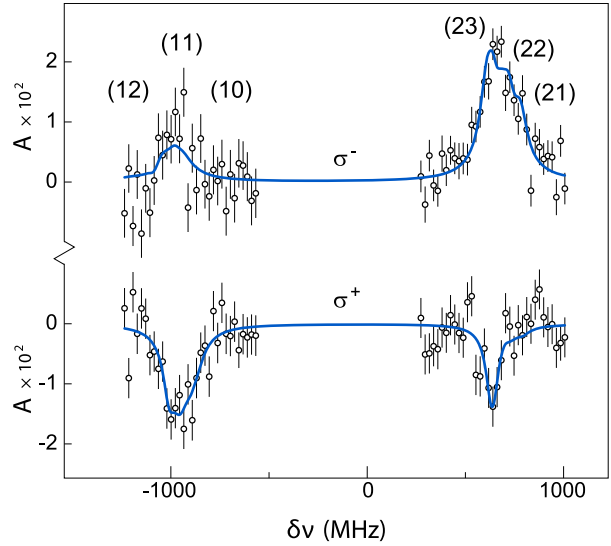}
			\caption{$\beta$-asymmetry spectra of $\sigma^\mp$ polarized $^{33}$Mg\,\textsc{ii} ($I^\pi=3/2^-$) fitted simultaneously with the realistic polarization function discussed in the text.}
			\label{fig2}
		\end{minipage}
	\end{center}
\end{figure}

\begin{figure}[t]
	\begin{center}
		\begin{minipage}[b]{0.45\textwidth}
			\includegraphics[angle=0,width=\linewidth]{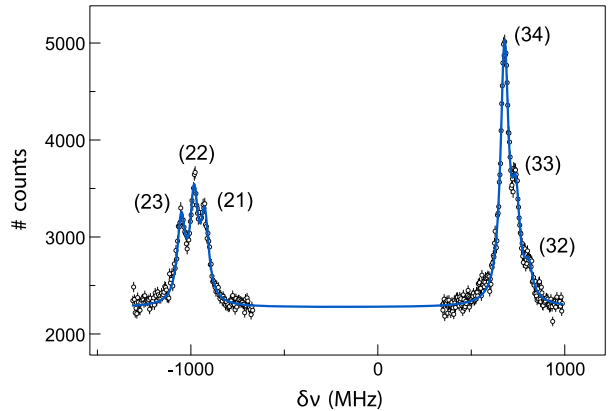}
			\caption{Fluorescence spectrum of $^{25}$Mg\,\textsc{ii} ($I^\pi=5/2^+$) fitted with asymmetric Voigt profiles from Ref.~\cite{Yord}.}
			\label{fig3}
		\end{minipage}
	\end{center}
\end{figure}

The hyperfine-structure of $^{29}$Mg in Fig.~\ref{fig1} is partly resolved. In the inset (a) the data are fitted with the realistic polarization function discussed in our previous work \cite{DTYordanov,Yordanov,Kowalska,Keim}. The magnetic hyperfine parameter in the $3p\;^2P_{3/2}$ state is substituted with the value reported in Ref.~\cite{Kowalska}, thus the relative position of the resonances is optimized by the variation of the quadrupole hyperfine parameter. In order to obtain a handle on the systematic uncertainty associated with the choice of fit function we have performed in the inset (b) a basic fit using three Lorentzian profiles. The resulting quadrupole hyperfine parameters in the two cases are $-23(2)$~MHz and $-15(2)$~MHz, respectively. As a final result in Tab.~\ref{tab1} we quote the mean value of the two, with half the difference being adopted as the associated systematic uncertainty and being added in quadrature to the statistical ones. The spectrum of $^{33}$Mg in Fig.~\ref{fig2}, and its fit, have been previously discussed in connection with the corresponding NMR measurement \cite{DTYordanov}, thus fixing the nuclear spin and the sign of the associated magnetic moment. With the magnetic splitting tied to the measured $g$ factor one extracts the quadrupole hyperfine parameter shown in Tab.~\ref{tab1}. Both isotopes have been referenced to $^{25}$Mg whose hyperfine structure is shown in Fig.~\ref{fig3}. The fit therein, using asymmetric Voigt profiles \cite{Yord}, is aided by fixing the magnetic hyperfine parameter of the $3s\;^2S_{1/2}$ state to the precise value from Ref.~\cite{Itano}. Thus, one determines a ratio of $A(3p\;^2P_{3/2})/A(3s\;^2S_{1/2})=0.032(1)$ and the $B$ factor presented in the table. The quadrupole moments in Tab.~\ref{tab1} are calculated proportionally to the know value of $^{25}$Mg \cite{Sundholm}.

\begin{table}[t]
\caption{\label{tab1}
Hyperfine $B$ parameters and quadrupole moments from this work, compared with shell-model calculations.
}
\begin{tabular}{ccccc}
\hline
						&	$I^\pi$		& $B~{(3p\;^2P_{3/2})}$~(MHz)	&	$Q$~(mb)											&	$Q_\text{theory}$~(mb)	\\
\hline
$^{25}$Mg		& $5/2^+$	 &	$+24(2)$										& $+199.4(20)$\footnotemark[1]	& $+217$\footnotemark[2]	\\
$^{29}$Mg		& $3/2^+$	 &	$-19(5)$										& $-158(44)$										& $-110$\footnotemark[2]	\\
$^{33}$Mg		& $3/2^-$	 &	$+16(11)$										& $+134(92)$										& $+157$\footnotemark[3]	\\
\hline
\end{tabular}
\footnotetext[1]{Reference quadrupole moment \cite{Sundholm}}
\footnotetext[2]{Shell-model calculations using the USDB Hamiltonian \cite{Brown}}
\footnotetext[3]{The two particle-hole value from Ref.~\cite{DTYordanov}}
\end{table}

\section{Discussion}

Shell-model calculations have been carried out as following. For $^\text{25, 29}$Mg we used the universal $sd$ Hamiltonian USDB \cite{Brown} and the code NuShellX~@~MSU \cite{BrownRae} while the calculated quadrupole moment of $^{33}$Mg is adopted from Ref.~\cite{DTYordanov} where the code Oxbash was used \cite{BABr} and the $sd\text{-}pf$ interaction from Ref.~\cite{Nummela}. The agreement with experiment for all three cases is generally good. In fairness to the reader there is no specific addition to make to previous discussions, the main reason being not the limited precision, but mostly nature itself. In the case of $^{33}$Mg the configurations with one and two particle-hole excitations produce nearly identical quadrupole moments \cite{DTYordanov}. The state with no cross-shell excitations can indeed be excluded on the basis of this work, albeit it is easily done so on the basis of the magnetic moment alone. Both theory and experiment are consistent on the $sd$ cases of $^\text{25, 29}$Mg, which have been clearly dissociated from the island of inversion by former studies.

\section{Conclusions}

In summary, we have determined the quadrupole moments of $^{29}$Mg and $^{33}$Mg with an uncertainty of about 30\% and 70\%, respectively. To our knowledge, these are the first quadrupole moments derived from $\beta$-asymmetry detected hyperfine structure. It should be noted that their accuracy is affected by the unresolved levels of the excited atomic state and by the experimental procedures which were not optimized for higher resolution. Higher precision and accuracy are in principle possible in a dedicated experiment. Comparison with shell-model calculations does not contradict the well-established picture of light nuclei in the region.

\begin{acknowledgments}

This work has been supported by the German Federal Ministry for Education and Research under contract no.~06MZ175 and 06MZ215, the Helmholtz Association (VH-NG-037), the P6-EURONS (RII3-CT-2004-506065), the IUAP project P5/07 of OSCT Belgium, the FWO-Vlaanderen and the Marie Curie IEF program (MEIF-CT-2006-042114). We thank the ISOLDE technical group for their professional assistance.

\end{acknowledgments}

\end{document}